# Might I Get Pwned:
## A Second Generation Compromised Credential Checking Service

Bijeeta Pal[†], Mazharul Islam[‡], Marina Sanusi[†], Nick Sullivan[*], Luke Valenta[*],
Tara Whalen[*], Christopher Wood[*], Thomas Ristenpart[††], Rahul Chattejee[‡]

[†]*Cornell University,*  [‡]*University of Wisconsin–Madison,*  [*]*Cloudflare,*  [††]*Cornell Tech*

## Abstract

Credential stuffing attacks use stolen passwords to log into victim accounts. To defend against these attacks, recently deployed compromised credential checking (C3) services provide APIs that help users and companies check whether a username, password pair is exposed. These services however only check if the exact password is leaked, and therefore do not mitigate credential tweaking attacks — attempts to compromise a user account with variants of a user's leaked passwords. Recent work has shown credential tweaking attacks can compromise accounts quite effectively even when the credential stuffing countermeasures are in place.

We initiate work on C3 services that protect users from credential tweaking attacks. The core underlying challenge is how to identify passwords that are similar to their leaked passwords while preserving honest clients' privacy and also preventing malicious clients from extracting breach data from the service. We formalize the problem and explore ways to measure password similarity that balance efficacy, performance, and security. Based on this study, we design "Might I Get Pwned" (MIGP), a new kind of breach alerting service. Our simulations show that MIGP reduces the efficacy of state-of-the-art 1000-guess credential tweaking attacks by 94%. MIGP preserves user privacy and limits potential exposure of sensitive breach entries. We show that the protocol is fast, with response time close to existing C3 services. We worked with Cloudflare to deploy MIGP in practice.

## 1 Introduction

Users often pick the same or similar passwords across multiple web services [22, 42, 54]. Attackers therefore compromise user accounts using passwords leaked from other websites. This is known as a credential stuffing attack [25]. In response, practitioners have set up third-party services such as Have I Been Pwned (HIBP) [37, 48], Google Password Checkup (GPC) [44, 47], and Microsoft Password Monitor [33] that provide APIs to check if a user's password has been exposed in known breaches. Such breach-alerting services, also called compromised credential checking (C3) services [37], help prevent credential stuffing attacks by alerting users to change their passwords.

Existing C3 services, however, can leave users vulnerable to credential tweaking attacks [22, 41, 51] in which attackers guess variants (tweaks) of a user's leaked password(s). Pal et al. [41] estimate that such a credential tweaking attacker can compromise 16% of user accounts that appear in a breach in less than a thousand guesses, despite the use of a C3 service.

We therefore initiate exploration of C3 services that help warn users about passwords similar to the ones that have appeared in a breach. We design "Might I Get Pwned" (MIGP, the name is a tribute to the first-ever C3 service, HIBP). In MIGP, a server holds a breach dataset D containing a set of username, password pairs $(u_i, \tilde{w}_i)$. A client can query MIGP with a username, password pair $(u, w)$, and learns if there exists $(u, \tilde{w}) \in D$ such that $w = \tilde{w}$ or $w$ is similar to $\tilde{w}$. To realize such a service, we must (1) determine an effective way of measuring password similarity, that (2) works well with a privacy-preserving cryptographic protocol, and that (3) resists malicious clients that try to extract entries from D.

Ideally, we want our similarity measure to help warn users if their password $w$ is vulnerable to online credential tweaking attacks. These attacks [22, 41, 51] take as input a breached password $\tilde{w}$ and generate an ordered list of guesses. Therefore, a good starting point for defining similarity is to call $w$ similar to $\tilde{w}$ should $w$ appear early in the guess list generated by a state-of-the-art credential tweaking attack. Such a generative approach also works well with simple extensions to existing cryptographic private membership test (PMT)-based protocol [37, 47]. A PMT allows a client to learn if $(u, \tilde{w}) \in D$ without revealing it to the server. To extend, we can have the server insert $n$ variants of each breached password into D and we can allow clients to generate $m$ variants and repeat the PMT for each of them. The PMT can be designed to reveal, upon a match, whether a password matches the original password or a variant.

To concretize this approach requires understanding how to efficiently generate effective variants. Existing credential tweaking attack algorithms are computationally expensive to run [41, 51], and it is unclear, apriori, what are good values for *m* and *n*. We use empiricism to explore different techniques for enumerating variants and show via simulations how these techniques help protect against credential tweaking attacks. We start with the deep learning [41] and mangling rules techniques [22] pioneered in prior works on credential tweaking. We also suggest a new, simple-to-implement generative approach that uses an empirically-derived weighted edit distance to rank mangling rules. We show via simulation that our new approach with $m = 10$ and $n = 10$ reduces credential tweaking attack success rate by 94% compared to using only exact-checking, where the attacker uses a thousand guesses and adapts to the breach alerting service being used.

Another challenge for MIGP services is *breach extraction attacks*. C3 services could contain breach data that is not publicly available. Most C3 services provide public APIs, which malicious clients can abuse to learn a user's breached passwords by querying the service with a sequence of likely passwords. MIGP services may make such extraction attacks faster, because, intuitively, finding one of many variants of the target password would also reduce the search space.

We formalize this new breach extraction attack setting and show that optimal strategies for an attacker are NP-hard to compute. Nevertheless, attackers can use heuristic approximations. We evaluate such heuristics empirically for various values of *n* and *m*. Our simulation shows that an attacker can compromise $2.8\times$ more user accounts in 1,000 guesses for server-only variant generation ($n = 100$) than the best attack against a traditional exact-checking service. Allowing a hybrid of client-side ($m > 0$) and server-side variant generation leads to even more effective attacks.

We therefore propose a blocklisting strategy to reduce breach extraction success rates: remove (blocklist) most popular passwords and their variants. Users should be warned to avoid such easy-to-guess passwords whether or not they appear in a breach. Blocklisting the most common $10^4$ passwords can reduce the success rate of the best-known breach extraction attack against a MIGP service to below the success rate possible against currently deployed C3 services.

We implement a prototype of MIGP with 1.14 billion breached username, password pairs, and show that online computation work for the server is small, client-side latency is comparable to existing C3 services (500 ms), and certain parameter regimes allow bandwidth required to be less than 1.43 MB. We further empirically explore different trade-offs in performance and security for client-side, server-side variant and hybrid generations for MIGP to help practitioners decide which approach to use. All this helped educate our deployment of MIGP in collaboration with Cloudflare, a major CDN and security service provider [9]. It is now in production use in their web application firewall product to notify login servers about potential attacks.

**Contributions.** The main contributions of this paper are:

- We initiate exploration of similarity-aware C3 services and present the design of MIGP, which allows checking if a password is vulnerable to credential tweaking attack without revealing it to the MIGP server.
- We empirically evaluate the effectiveness of different similarity measures to mitigate credential tweaking attacks.
- We analytically and empirically analyze the threat of breach extraction attacks, in which malicious clients attempt to extract credentials from a C3 service. We discuss multiple approaches to mitigate this threat, including a new popular-password blocklisting mechanism.
- We report on an initial prototype of MIGP and show its practicality by deploying at Cloudflare.

## 2 Background and Prior Work

**Credential stuffing attacks and defenses.** Billions of passwords are available online as a result of compromises [25, 48]. As users often choose the same or similar password for different web services [22, 42, 51], attackers use these leaked data for credential stuffing attacks. As a prevention measure, C3 services have been adopted in client browsers [44], in password managers [2], and by login server backends to proactively check user credentials. Existing C3 services include Have I Been Pwned (HIBP) [48], Google Password Checkup [44], Enzoic [4], and the recently introduced Microsoft Password Monitor [33]. HIBP [48] has publicly documented APIs to check if a username or password is in a breach. Several password managers such as 1Password and LastPass and browsers such as Firefox are using HIBP to warn users about their leaked passwords. This may result in false positives since common passwords will always be flagged.[1]

Google Password Checkup (GPC) [47], released as a Chrome-extension in 2019 [44] and later integrated into Chrome, checks if a username, password pair is present in the leak, leading to fewer false positives compared to HIBP. The Chrome password manager uses GPC to check all of a user's website credentials to determine if they are in a known breach, but does not flag passwords that are similar to ones in breaches. Li et al. [37] formalized the security requirements of C3 systems in an honest-but-curious server setting and proposed a protocol that we use to build MIGP in this paper.

The state-of-the-art C3 protocol proposed in [37, 47] now deployed by GPC handles a large scale of breach datasets using *bucketization*. To check a username, password pair $(u, w)$, the client sends a bucket identifier *j* which is the first 16 bits of the cryptographic hash of *u* (smaller hash prefix helps preserve the privacy of the username). In parallel, the client and server perform a private membership test (PMT) protocol to securely determine if $(u, w)$ is in the bucket containing the set of all $(u_i, \tilde{w}_i)$ with the same username hash prefix. The PMT protocol is built using the efficient oblivious PRF (OPRF) protocol, 2HashDH [32], though a recently proposed partially oblivious PRF 3HashSDHI may be used to slightly improve security [49]. A more recent service, Microsoft Password Monitor [33], uses homomorphic encryption (HE) to compute the PMT, but reveals the username completely to the server.

---
[1] We found flagging based on only passwords will raise 29% false alarms, and based on only usernames will raise 36% false alarms to users whose passwords might not be vulnerable to a credential tweaking attack.

To prevent users from reusing their password across web services, Wang and Reiter [52, 53] proposed protocols to check if a user is using the same password in multiple participating web services. The efficacy of this protocol relies on the coordination of the web services, making it harder to deploy. Moreover, as we show in Section 6, the PMT protocols used in their work would not scale to billions of username, password pairs without sacrificing the privacy of the username. Wang and Reiter also mention that their protocol can be extended to check for similar passwords across multiple web services [53], but did not provide details on how to do so.

**Credential tweaking attacks and defenses.** Currently deployed C3 services cannot warn users about a password unless the exact password is present in the breach. For example, a minor variation, such as adding "7" to the end of the compromised password "yhTgi456", won't be detected by the C3 service. Users often pick similar passwords while resetting their passwords on a web service [54] or when picking passwords for different web services [22]. These passwords are vulnerable to credential tweaking attacks [22, 41, 51], where the attacker tries different variations of the leaked password.

Wang et al. [51] and Das et al. [22] used human-curated rules to generate guesses for a credential tweaking attack. Subsequently, Pal et al. [41] took a data-driven, machine-learning approach to build similarity models for passwords from the same user. They trained a sequence-to-sequence [46] style neural network model (pass2path) that outputs similar passwords given an input password. This is now the best-known attack, with simulation showing that a pass2path-based attack can compromise 16% of accounts of users that appeared in a breach using at most 1,000 guesses, despite the use of a C3 service as a credential stuffing countermeasure. Pal et al. also showed in a case study that over a thousand accounts at Cornell University were at the time vulnerable to credential tweaking attacks, showcasing their practical risk.

Pal et al. proposed a potential defense: a personalized password strength meter (PPSM) which considers the strength of a selected password based on its similarity to the user's other passwords. But they do not offer a way to utilize PPSMs in the context of a privacy-preserving C3 service, and left building similarity-checking C3 services as an open question.

## 3 Overview of MIGP

In this paper, we build a similarity-aware C3 service, called Might I Get Pwned (MIGP). MIGP generalizes existing C3 services to add new features that can warn users about passwords that may be vulnerable to credential tweaking attacks.

**Service architecture and functionality.** The MIGP server will have a breach dataset D, containing a set of |D| username, password pairs $\{(u_1, w_1), \ldots, (u_{|D|}, w_{|D|})\}$ where each $u_i \in U$ is a username and each $w_i \in W$ is a password. The sets $U$ and $W$ consist of all possible user-chosen usernames and passwords. A MIGP client can query the MIGP server with a username, password pair $(u, w)$ to learn if there exists a $(u, \tilde{w}) \in D$ such that $w = \tilde{w}$ or $w$ is *similar* to $\tilde{w}$. The MIGP server, therefore, returns "match" if $w = \tilde{w}$, returns "similar" if $w$ is similar to $\tilde{w}$, and returns "none" otherwise.

A MIGP client can be, for example, a user's browser, their password manager, an authentication service, or, as in our Cloudflare deployment, a web application firewall that wants to use breach alerting to help secure user accounts. We will use as a running example the user's browser as client, and discuss other deployment settings in Section 7.

Like existing C3 services, MIGP should scale to millions of requests a day with billions of username, password pairs in its database. We propose various techniques to make MIGP fast and practical, like offline processing the breach data to speed up online queries and rate-limiting clients using verifiable delay functions rather than slow hashing (Appendix F).

**Threat model.** In our threat model, we consider two distinct threats: (1) an honest-but-curious server trying to learn about a user's queried password, and (2) a malicious client querying the MIGP server to retrieve other users' breached passwords.

We assume the MIGP server is honest-but-curious: it doesn't deviate from the protocol but observes the protocol in an attempt to glean information from the user queries. Technically, we note that our MIGP protocol is in fact one-sided simulatable [29], a model which allows the server to behave maliciously. But for practical purposes, a malicious server can misguide a user by returning a wrong bucket of passwords and falsely reporting the user's vulnerable password as safe (i.e., an input-switching attack). Regular audits and other monitoring techniques may be useful mitigations. We are not aware of any other active attacks and will focus on the honest-but-curious server setting hereafter.

Ideally, we would like the MIGP server (or any C3 server) to learn nothing about the queried usernames or passwords. However, building a practical solution that achieves this requirement is hard given the huge scale of D with billions of credentials. The state-of-the-art protocols in existing C3 services reveal some bits of information about the username to allow partitioning D into smaller buckets on which a private membership test (PMT) protocol can be efficiently executed. Looking ahead, MIGP will extend this approach to perform a private similarity test (PST) over the bucket.

Clients of MIGP can be malicious. In particular, they might mount a guessing attack in an attempt to extract username, password pairs from D. We call this a *breach extraction attack*. These are a concern when the breach database D contains data from leaks that are not yet publicly available. In turn, learning a user's (leaked) password can help the attacker compromise that user's accounts on other web services through credential stuffing and tweaking attacks. Prior work did not empirically analyze this threat for exact-check C3 services, but they did include anti-abuse countermeasures such as requiring computationally intensive slow hashing to complete a query [47]. This threat is particularly concerning for MIGP as clever attacks may exploit similarity.

**Unsatisfactory approaches.** The core of MIGP is a password similarity metric. While there are a number of ways to compute password similarity, few can preserve the privacy of the queried password. For example, Pal et al. [41] design password embedding models that map passwords to a vector space; distance in the space captures similarity. Using password embeddings directly (e.g., the client sending a password embedding to the server) is unsafe as it might reveal the underlying passwords.

One can instead build a MIGP service by combining a password embedding with a secure two-party computation (2PC) protocol that privately computes the dot product and threshold comparison. However, even state-of-the-art 2PC protocols for computing dot products [34] are not yet efficient enough to be used in our setting (which will require computing thousands of such dot products per query). We estimate, based on a prototype implementation using a 2PC library named Crypten [34], that a single client query would take 16 seconds (without network latency) to complete private dot product and comparison (Appendix A). Other approaches that rely on existing secure two-party computation protocols, such as computing a weighted edit distance between passwords, will similarly fall short of our performance requirements.

**Generative models for password similarity.** We instead use a generative approach to measure similarity, which will enable more efficient privacy-preserving protocols. We consider generative approaches that either start with a breached password or with a client's password. For the former, let $\tau_n: W \mapsto W^n$ be a function that generates $n$ passwords that are likely to be chosen by a user, given one of their other breached passwords. Thus, a client password $w$ and breached password $\tilde{w}$ are similar if $w \in \tau_n(\tilde{w})$. Here, we assumed $w \notin \tau_n(w)$ for all $w \in W$. For the second approach, an inverse generative model, say $\tilde{\tau}_m$, generates $m$ variants given a client's password; we declare a password similar to a variant if $\tilde{w} \in \tilde{\tau}_m(w)$. Because similarity is not necessarily symmetric, it can be that $\tau_n \neq \tilde{\tau}_m$. Looking ahead, we can use the model $\tau_n$ to generate likely variants at the server given a breach dataset, while we can use $\tilde{\tau}_m$ to generate variants at the client. We will also explore a hybrid approach that combines the two, in which case we consider a client's password $w$ similar to a variant $\tilde{w}$, if $\bigl(\{w\} \cup \tilde{\tau}_m(w)\bigr) \cap \bigl(\{\tilde{w}\} \cup \tau_n(\tilde{w})\bigr) \neq \varnothing$ and $\tilde{w} \neq w$. A big question we will tackle is how to best instantiate $\tau_n$ and $\tilde{\tau}_m$.

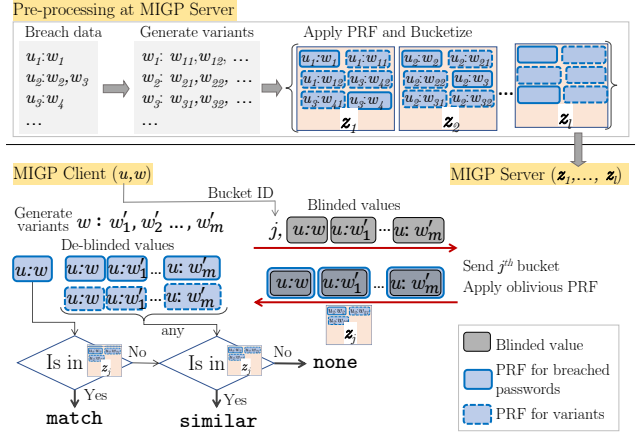

Figure 1: MIGP Protocol for checking if a queried password $w$ is similar to a password present in breach data. Cryptographic details of the protocol are given in Fig. 2.

## 3.1 MIGP protocol

MIGP builds off first-generation C3 designs, specifically, the identity-based bucketization (IDB) protocols due to Li et al. [37] and Thomas et al. [47]. At a high level, the IDB protocol splits the leaked credential database into several buckets based on truncated hashes of usernames. The client reveals the bucket identifier to the service, and then performs an OPRF-based private membership test (PMT) protocol over that bucket to check for equality.

In Fig. 1, we provide an overview on how to extend IDB to allow the client to check for similar passwords. We augment the server's breach data with variants of each breached password using $\tau_n$. The client queries the server using the IDB protocol with the user password and checks if it succeeds. The client can also generate variants, via $\tilde{\tau}_m$. There are nuanced security and computation trade-offs for this approach, which we will discuss at the end of this section. For now, we assume the client and the server both generate $m$ and $n$ variants using $\tilde{\tau}_m$ and $\tau_n$ functions. Setting $m = 0$ and $n = 0$, reduces MIGP functionality to existing exact-checking C3 services, such as IDB. The cryptographic details of MIGP protocol, which fits our security requirements, is given in Fig. 2.

**Pre-processing.** The underlying IDB protocol uses a specialized oblivious PRF construction. Briefly, the PRF takes as input a username $u$, a password $w$, and a secret key $\kappa$ and is defined as $F_\kappa(u\|w) = \mathsf{H}_2(u\|w, \mathsf{H}_1(u\|w)^\kappa)$. This is the same as the 2HashDH construction due to Jarecki et al. [32]. Here $\mathsf{H}_1$ maps onto an elliptic curve group $\mathbb{G}$ (with group operation written multiplicatively) where the decisional Diffie-Hellman (DDH) problem is hard; and $\mathsf{H}_2 : \{0,1\}^* \times \mathbb{G} \mapsto \{0,1\}^\ell$ maps a binary string and a group element to an $\ell$-bit string. At least one of the two hash functions used should be computationally expensive (for the client) to ensure rate limiting and abuse prevention on the client side. We explore trade-offs on how to choose the hash functions in Section 7 and Appendix F.

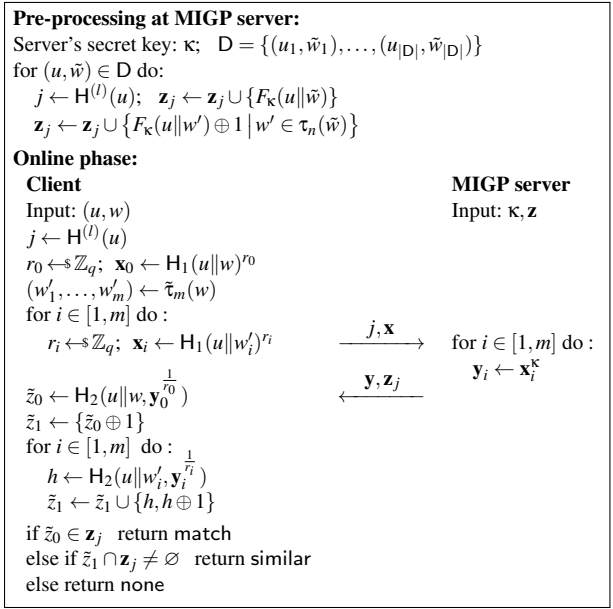

```
Pre-processing at MIGP server:
Server's secret key: κ;  D = {(u_1, w̃_1), ..., (u_{|D|}, w̃_{|D|})}
for (u, w̃) ∈ D do:
    j ← H^{(l)}(u);   z_j ← z_j ∪ {F_κ(u‖w̃)}
    z_j ← z_j ∪ {F_κ(u‖w') ⊕ 1 | w' ∈ τ_n(w̃)}
Online phase:
  Client                                          MIGP server
  Input: (u, w)                                   Input: κ, z
  j ← H^{(l)}(u)
  r_0 ←$ Z_q;   x_0 ← H_1(u‖w)^{r_0}
  (w'_1, ..., w'_m) ← τ̃_m(w)
  for i ∈ [1, m] do:
      r_i ←$ Z_q;   x_i ← H_1(u‖w'_i)^{r_i}     --j, x-->     for i ∈ [1, m] do :
                                                                  y_i ← x_i^κ
                                                <--y, z_j--
  z̃_0 ← H_2(u‖w, y_0^{1/r_0})
  z̃_1 ← {z̃_0 ⊕ 1}
  for i ∈ [1, m] do :
      h ← H_2(u‖w'_i, y_i^{1/r_i})
      z̃_1 ← z̃_1 ∪ {h, h ⊕ 1}
  if z̃_0 ∈ z_j  return match
  else if z̃_1 ∩ z_j ≠ ∅   return similar
  else return none
```

Figure 2: Protocol for checking if a password similar to the user's password ($w$) is present in the leaked data (D).

The server chooses $\kappa$ and applies $F_\kappa$ to all the username, password pairs in the breach. These are stored separate "buckets", identified by the $l$-bit prefix of a cryptographic hash of the username, denoted $H^{(l)}(u)$. As we want the client to find out if the queried password is similar to one stored by the server, we use two PRF functions: The server stores $F_\kappa(u\|w)$ (shown in thick blue border boxes in Fig. 1) corresponding to the leaked credential $(u, w)$, and $F'_\kappa(u\|w') = F_\kappa(u\|w') \oplus 1$ for $w' \in \tau_n(w)$ corresponding to the password variants, which is represented by the dashed blue boxes in Fig. 1. The last bit of the PRF of similar passwords is flipped to differentiate it from the original leaked password.

**Online computation.** MIGP client, on input a user id $u$ and password $w$, calculates the ID of the bucket to query based on the username, $j = H^{(l)}(u)$. Then the user generates $m$ variants of their password $w$ based on $\tilde{\tau}_m$. The client "blinds" the passwords and their variants, sending to the server $H_1(u\|w)^{r_0}, H_1(u\|w'_1)^{r_1}, \ldots, H_1(u\|w'_m)^{r_m}$ for random values $r_0, \ldots, r_m \in \mathbb{Z}_q$. Blinding ensures that the MIGP server does not learn anything about the query (beyond $j$).

The server raises each of the blinded values to the secret key $\kappa$, and sends these back to the client, along with the bucket $z_j$. The client can deblind the values to finish computing the PRF on all $m+1$ values. Then it checks if $F_\kappa(u\|w)$ is present in the bucket, and if so, it learns that $(u, w)$ is in the leaked data, outputting match. If not, the client checks if any other computed PRF values $F_\kappa(u\|w'_i)$, or those values with last bit flipped $F_\kappa(u\|w'_i) \oplus 1$, or $F_\kappa(u\|w) \oplus 1$ is in the bucket. If any are found, then the client learns that $(u, w)$ is similar to a $(u, w')$ found in the password breach, outputting similar. Otherwise, it outputs none.

## 3.2 Server- vs. client-side variant generation

Based on the values of the parameters $n$ and $m$, MIGP protocol can allow generating variants only on the client-side ($n = 0$), only on the server-side ($m = 0$), or a mix of both. By allowing variants only on the server side, the existing IDB protocol can be easily adopted, making it simpler to implement. However, the server database expands by $n$ times, requiring more disk space and more bandwidth due to larger buckets.

In the case of client-side generation of variants, no change on the server is required. The variants can be batched together in a single API query to the server, saving network round trips and bandwidth. (Note, the client only needs to download the matching bucket from the server once per username.) Moreover, in this approach, the client will have more control over the variations. It can use inputs from the user, such as their other passwords, to generate personalized variants that are likely to be used as passwords by that particular user. Such personalization was shown to be useful for correcting password typos [21] and could be also useful for MIGP.

Although the client-side generation of password variants has some benefits, it also suffers from some key limitations. First, existing C3 services have rate-limiting measures, like slow hashing, to prevent malicious clients from extracting the breach data by repeatedly calling the APIs with different password guesses [44]. This would make checking multiple variations of a password too expensive to be practical. To make things faster, the server could allow batching all queries into one request and reduce the client-side computation. But there is a key security issue with this approach: as the OPRF protocol blinds queried values, the server cannot differentiate if a query contains a set of variants of a password or completely different passwords. This can be exploited by a malicious client to obtain a factor of $n$ improvement in breach extraction attack efficacy (Section 5.1). Zero-knowledge proofs [28] could be used, in theory, to prevent a malicious client from checking arbitrary passwords, but it remains an open question whether they can be made practical in this setting. We leave finding an efficient solution to this problem for future work.

In the hybrid approach, the client generates $m$ personalized variations, possibly based on their other passwords or personal information, and the server also stores $n$ variations of each breached password. Such a protocol with appropriate client and server-side generation functions can increase the protection against credential tweaking attacks to the equivalent of generating $n \times m$ variants on the server or the client side (as we show in Section 4). The hybrid approach can also reduce the storage cost on the MIGP server, reduce bandwidth cost due to smaller buckets, and lower the advantage gained in breach extraction attack by allowing a smaller number of guesses per malicious query.

In subsequent sections, we explore the performance, security, and efficacy implications of different choices of $m$, $n$, along with how to build practical generative models $\tilde{\tau}_m$, $\tau_n$.

## 4 Efficacy of Different Similarity Measures

We explore different measures of password similarity using generative models that enumerate the most likely variants of a given password. Though the client-side and server-side models can be different, we cannot learn two different models due to the limitation of our dataset (as we explain below). Thus we will focus on building a single generative model $\tau$ that will be used both on the client and server side; we will show even this simple approach already performs well.

In particular, we compare different similarity measures $\tau$ based on efficacy at protecting from credential tweaking attacks, computational performance, and security against breach extraction attacks. We focus on the first two in this section, and discuss the third in Section 5.

### 4.1 Similarity measures

As we focus on generative similarity metrics, any credential tweaking attack can be repurposed to be a similarity metric. We, therefore, start with the attack algorithms proposed in Das et al. [22] and Pal et al. [41]. We denote these by Das and P2P, respectively. We also created more efficient and effective variants of these methods, named Das-R and wEdit, as we discussed below. Each method takes as input a password $w$ and outputs an ordered list of similar passwords.

We also compare the generative methods to the embedding-based similarity measure, PPSM, proposed in [41]. Although existing PST protocols suitable for use with PPSM are not fast enough for use in practice (as discussed in Section 2), we still discuss them here should PST protocols become more suitable for deployment in the future.

**Das.** Das et al. [22] were the first to show that users select similar passwords across multiple websites, and that it is easy to guess a user's password given one of their other passwords. They, given a password $w$, use a set of hand-crafted tweaks to generate similar passwords. We refer to this approach of generating similar passwords as Das.

**Das-R.** We observed that ordering of the tweaks used in Das is not effective for smaller $n$. So we reorder the set of tweaks based on the frequency with which these tweaks are used by users in our dataset (Section 4.2). We show the reordering significantly improves the efficacy of the rules when considering smaller numbers of variants ($\leq 10$). We call this similarity measure Das-R. The reordered rules are given in Appendix B. Not all tweaks apply to all passwords, in which case we continue applying further tweaks until we obtain $n$ variants.

**P2P.** While Das et al. used hand-crafted tweaks for generating variations, Pal et al. [41] used a neural network model, called pass2path (P2P), to learn the tweaks a user is likely to make to their passwords. This resulted in the most damaging credential tweaking attack to date, outperforming prior works, such as [22] and [51]. We refer to this approach as P2P. While P2P is quite effective at capturing password similarity, it is slow and expensive (even with GPUs) to compute.

**wEdit.** Finally, we explore automatically deriving a ranked list of tweaks that can be applied to a password to obtain variants. Although tweaks have long been used in password cracking systems (e.g., [13]), here the goal is different — finding variants likely to be chosen by a user and that are vulnerable to credential tweaking attacks.

Following the definitions in [41], we define a *unit transformation* as a specific edit to be applied to the input password $w$. A unit transformation is defined by a tuple $(e, c, l)$ where $e$ specifies the edit type as one of insert, delete, or substitute; $c$ denotes the character to be inserted or substituted ($c = \perp$ for deletion); and $l$ is the location for the edit. The location is length-invariant, representing the distance from the first character by positive numbers and from the last character by negative numbers; we use the smaller of the two distances and break ties using the distance from the start of a password. For example, (insert, '0', $-1$) specifies adding the character '0' to the end of a password, and (substitute, 'a', 2) specifies replacing the second character with a lowercase letter 'a'.

Given a pair of passwords $(w, w')$, we can calculate the shortest sequence of unit transformations to generate $w'$ from $w$. We refer to this as the *transformation path*. The computation can be done using standard edit distance algorithms. We use the keypress representation of the passwords $w, w'$ as defined in [20], which includes special characters such as shift and caps lock.

Given a breach dataset containing multiple passwords associated with the same user accounts, we compute transformation paths for every pair of passwords belonging to the same user. Then we create a ranked list of transformation paths based on how many pairs of passwords it explains. To generate variants of a password $w$, apply the transformation paths one at a time, in decreasing frequency order, skipping if it is not applicable. We stop if we have generated $n$ variants. Note that wEdit contains a much more exhaustive list of tweaks (transformation paths) compared to Das-R. However, wEdit is not sensitive to the input unlike the handcrafted rules in Das-R, which include rules like insert '3' if the last character of the word is '2'. (The rules for wEdit and Das-R are given in Appendix B.) Nevertheless, we will see below that they have similar efficacy in our context.

### 4.2 Breach dataset

To drive empirical evaluation of the five similarity approaches, we use a dataset containing a compilation of publicly available breaches on the Internet [17]. This dataset was also used in prior academic research work and industry reports, e.g., [25, 37, 41], and has been confirmed to contain real user accounts. The breach compilation dataset contains nearly 1.4 billion unique email, password pairs. We clean the dataset based on the procedure described in [41], such as removing passwords containing non-ASCII characters or longer than 30 characters (which affects only 0.3% of users). We merged usernames based on the mixed-method from [41] and removed users having more than 1,000 passwords. The resulting dataset D consists of 1.3 billion unique username-password pairs from 908 million unique users (Fig. 3). More details about the dataset can be found in [41].

| # | D | S | T | $S_1$ | $S_2$ |
|---|---|---|---|---|---|
| Users | 908 | 760 | 230 | 380 | 380 |
| Passwords | 438 | 373 | 119 | 210 | 210 |
| Unique user-pw pairs | 1,147 | 918 | 229 | 459 | 459 |
| Total user-pw pairs | 1,317 | 1,069 | 248 | 535 | 535 |

Figure 3: Number of unique users, passwords, and username, password pairs (in millions) in the entire dataset D, breach dataset S, and test dataset T. $S_1$ and $S_2$ are two equal partition of S. Total number of username, password pairs with duplicates shown in the last row.

For our simulations, we partitioned D into two: a larger split (80%) simulates the leaked dataset S, which we further divide into two equal sets $S_1$ and $S_2$ with no common users between them; and the remaining dataset (20%) is used as the testing dataset T. In Fig. 3, we report some statistics on the dataset splits. S and T consist of 760 and 230 million unique usernames, respectively. About 82 million usernames are present in the intersection of T and S, implying that these users in the test dataset have at least one password in the simulated breach dataset. The number of users, passwords, and user-password pairs are similar for $S_1$ and $S_2$, as expected.

For the attack simulations in Section 4.3, we conservatively assume the attacker has access to more data than what is known to the MIGP service. That is, we provide the attacker with the entire leaked dataset S but train the similarity mechanism for MIGP only on $S_1$ (training is needed for Das-R, P2P, and wEdit). The test dataset can, therefore, be considered a list of users' current passwords on some target websites for which the attacker wants to gain illicit account access. The test dataset is neither accessible to the attacker nor to the similarity mechanisms that we train.

### 4.3 Empirical efficacy evaluation

We examine the effectiveness of a similarity measure based on protection from credential tweaking attacks and impact on usability due to false warnings, which can cause user fatigue. To quantify this, we classify each pair of passwords belonging to the same user as *vulnerable* or *safe* based on whether or not they are vulnerable to credential tweaking attacks.

We pick password pairs $(w_1, w_2)$ belonging to the same user, such that $w_1$ is selected from $S_1$ and $w_2$ from T. Hence, both the attacker and service know the breached password $w_1$ corresponding to the target user and want to attack/protect the user's unknown (test) password $w_2$. We flag a pair vulnerable if $w_2$ can be guessed by pass2path [41] given $w_1$ in a thousand guesses. Otherwise, we flag the pair as safe. From all vulnerable pairs, we randomly sampled 10,000 pairs to measure the true positive rate (TPR) of a similarity measure $\tau$ as the fraction of vulnerable pairs that are flagged by it. Similarly, we randomly sampled 10,000 safe pairs and measured the false positive rate (FPR) of $\tau$ as the fraction of these pairs that are flagged by $\tau$, which burdens users with spurious warnings.

| Parameters | Similarity measures | % True positive | % False positive |
|---|---|---|---|
| $n = 10$ or $m = 10$ | Das | 33.2 | 0.6 |
| | Das-R | **52.6** | 0.0 |
| | P2P | 46.4 | 0.0 |
| | wEdit | 49.6 | 0.0 |
| $n = 100$ or $m = 100$ | Das | 46.9 | 2.2 |
| | Das-R | 63.5 | 0.2 |
| | P2P | **69.0** | 0.1 |
| | wEdit | **69.3** | 0.1 |
| $n = m = 10$ | Das-R | **89.9** | 2.9 |
| | wEdit | 75.2 | 2.2 |
| $n = m = 10$ (Greedy) | Das-R | **93.5** | 4.5 |
| | wEdit | 84.4 | 3.0 |
| $\theta = 0.83$ | PPSM | **67.9** | 2.0 |
| $\theta = 0.75$ | | 87.6 | 4.7 |
| $\theta = 0.5$ | | 99.1 | 14.0 |

Figure 4: True positive (ones vulnerable to 1,000-guess pass2path attack) and false positive (others) rates for different similarity measures, computed over 10,000 randomly sampled password pairs. The best performing measures are boldfaced.

The efficacy of a generative similarity measure can be different based on whether it is applied to the breached password (on the MIGP server, $\tau_n$) or to the queried password (on the client, $\tilde{\tau}_m$). For a pair of passwords $w, w'$ in the breach data, if we knew $w$ was used before $w'$, then we could train $\tau_n$ to generate edits that modify $w$ to $w'$ while $\tilde{\tau}_m$ consider variants of $w'$ leading to $w$. However, our training data does not contain such temporal ordering information. Therefore, as mentioned above, we use $\tilde{\tau}_m = \tau_n$, i.e., the variants are generated in the same way on the client and the server.

An orthogonal point is that for the hybrid case, in which both $m > 0$ and $n > 0$, better utility may come from considering simultaneously which rules should be used on the client and which ones should be used on the server. But the space of all possible $m \times n$ combined client-server rule sets is large, and we do not know how to search for optimal solutions efficiently. We used a greedy approach to understand the efficacy, but leave to future work developing better search techniques, and evaluating their potential for improving efficacy.

**Result.** Fig. 4 shows the performance of the similarity measures. As expected, increasing the number $n$ or $m$ of similar passwords improves the coverage against vulnerable pairs across all methods. However, that also increases the false positive rate, flagging safe passwords.

For MIGP where variants are only generated on the server (or on the client side), Das-R gives the maximum 52.6% coverage for $n$ or $m = 10$ tweaks among all the generative approaches. P2P and wEdit perform the best with $n$ or $m = 100$ with 69% coverage. PPSM gives high coverage against the attacks but also has a higher false-positive rate compared to the generative approaches. As the TPR of PPSM with reasonable FPR $\approx 2\%$ is lower than that of generative approaches (such as wEdit, $n$ or $m = 100$) and anyway does not lead to efficient protocol, we do not consider it further. Between wEdit and P2P, wEdit is drastically simpler to deploy and faster to run, requiring $4.5\times$ less pre-computation time (see Appendix A).

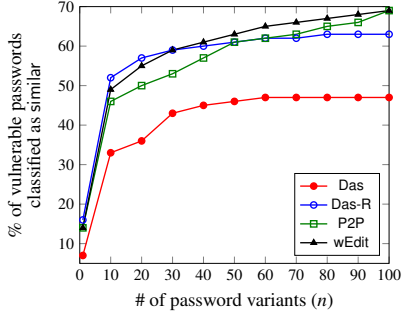

Figure 5: Percentage of vulnerable password pairs flagged by different similarity measures for varying $n$. The slopes of the graphs for all similarity measures decrease rapidly for $n > 10$.

The hybrid approach, where $n = 10$ variants are generated by the server and $m = 10$ are generated by the client both using Das-R rules, gives the best coverage to credential tweaking attacks, flagging 90% of vulnerable passwords at considerably low false flagging of safe passwords (2.9%). We also tried a greedy approach where we iteratively pick the tweaks on the client and the server that maximizes coverage of the tweaks until each side has $m$ and $n$ tweaks. This approach performs better at identifying vulnerable passwords, flagging 94% of them, but also has a high false positive rate (4.5%).

**Efficacy with increasing variants.** Fig. 5 examines how the efficacy of the four generative models varies by the increasing number of tweaks $n$ in server side variant checking. The results are the same for $m$ in the client-side MIGP. Das-R outperforms other techniques for $n \leq 30$. wEdit outperforms the other measures after that for $30 \leq n \leq 100$. It was surprising to us that the rule-based approaches (Das-R, wEdit) end up matching or exceeding the performance of the much more complex deep learning approach underlying P2P. This is because rule-based approaches can easily capture frequently seen variants, for low values of $n$. The deep learning approach works better for large $n$ by finding and ordering less frequently seen similarity relationships. For example, for $n = 10^3$, P2P outperforms wEdit by 4%.

Although increasing $n$ increases attack coverage, the slope of the curves decrease rapidly (Fig. 5). Therefore, the benefit of considering a higher $n$ value diminishes while increasing storage (only for server-side variant checking), computation, and bandwidth cost, as we see in Section 6.

Therefore in the rest of the paper, we use Das-R for $n$ or $m = 10$ or hybrid $n = m = 10$ and wEdit for $n$ or $m = 100$.

| Breach alerting method | $q=10$ | $q=100$ | $q=1000$ |
|---|---|---|---|
| Exact checking [37,47] | 10.1 | 13.4 | 16.3 |
| MIGP [Das-R, $n=10$ or $m=10$] | 2.8 | 5.0 | 7.9 |
| MIGP [wEdit, $n=100$ or $m=100$] | 1.9 | 3.0 | 5.2 |
| MIGP [Das-R, $n=10$ and $m=10$] | 0.6 | 1.0 | 1.4 |

Figure 6: Success rate of credential tweaking attacker in $q \in \{10, 100, 1000\}$ guesses, assuming that the attacker is aware of the breach alerting mechanism.

### 4.4 Adaptive credential tweaking attackers

We now measure the reduction in a credential tweaking attacker's success in breaking into a user account, should a MIGP service be deployed with one of the similarity measures discussed in Section 4.3. We compare against the baseline where an exact-checking C3 service such as [37, 47] is used. For the simulation, we adapt the best-known credential tweaking attack — pass2path [41] — to be aware of the MIGP service.

We conservatively assume that the attacker has access to the entire breach dataset $S$, while the MIGP service has access to the subset $S_1$. We sample 10,000 users from the test dataset $T$, who are also present in $S_1$ and have a password marked safe (not flagged as match or similar) by the service under consideration; this constitutes the target users for the attacker. With this user list, we can simulate the scenario where the service (the exact checking C3 service or the MIGP service) warned the user about their unsafe passwords on a target website and the user subsequently changed their password. Though not all users will abide by warnings, this setup allows us to compare the maximum security benefits of a service using similarity measures.

We consider an online attack setting, where too many incorrect password submissions should trigger an account lockout, resulting in attack failure. Thus the attacker has a query budget $q \leq 10^3$. We measure the fraction of user passwords the attacker can guess in $q$ attempts, assuming one of their other passwords is present in $S$. The attacker enumerates guesses by first generating candidates using pass2path, and skipping any that would be flagged by the service. The attacker can infer this themselves because we assume that the service's breach data and the similarity measure are known to the attacker.

As shown in Fig. 6, when only credential stuffing countermeasures are in place, such as using [37] or [47], the credential tweaking attacker can guess passwords of 10.1% of accounts using 10 guesses, which matches the performance reported in [41]. The hybrid MIGP reports the highest reduction in attack efficacy, 94% for $q = 10$. For the server or client-side MIGP service, the efficiency decreases to 1.9% when $n$ or $m = 100$ variations based on wEdit are used; a reduction of 81%. The attack accuracy decreases by nearly 78% and 68% for $q = 100$ and $q = 1,000$, respectively. Across the board, larger $n$ or $m$ gives better protection against the credential tweaking attacker, reducing the attack's efficacy.

```
MIGP(w'_1, ..., w'_m)                MIGPGuess(A, q):
─────────────────────              ─────────────────
q ← q − 1                           w* ←_p W
if q ≤ 0 then return none           w̃ ← A^MIGP
for i = 1 to m do                   if w̃ = w* return true
    if w'_i = w* then return (i, match)   else return false
    if w'_i ∈ τ(w*) then return (i, similar)
return none
```

Figure 7: An abstract breach extraction attack security game for a MIGP service parameterized by a number of MIGP protocol invocations $q$, a distribution of passwords $p$, a similarity model $\tau$, and a number of client-side variants allowed $m$.

## 5 Security Evaluation

A MIGP service allows clients to check whether a password similar to the queried one is present in the breach. Current C3 services, such as GPC [47], do not reveal any information about breach data unless a client queries the exact username, password pair. A natural question is: Will moving towards similar-aware C3 services degrade the confidentiality of the username, password pairs in the leaked database?

We formalize the abstract setting of a malicious client that has access to a similarity oracle, assuming it is cryptographically secure (we refer to this as the ideal functionality following parlance from the 2PC literature).

### 5.1 Breach extraction attacks

A MIGP service could be abused by malicious clients that seek to learn about user credentials. This is particularly concerning should a MIGP service have access to relatively new breaches that are not widely available to attackers, making the service a potential target for what we call a *breach extraction attack*. We model such attacks via the security game given in Fig. 7. In it, the adversary is given access to an oracle that implements the ideal functionality of a MIGP service. Note that the oracle is parameterized by a target password $w^*$ chosen by the game, the query budget $q$, and a similarity measure $\tau$. In each query, the adversary can send up to $m$ passwords, and each is checked against the target $w^*$ and its variants $\tau(w^*)$. Here we use $\tau$ for the server-side variants, but allow a malicious client to choose any $m$ passwords for the client-side variants. The goal of the attacker is to guess $w^*$ within the given query budget $q$.

Finding an optimal guessing strategy for breach extraction is NP-hard. (See Appendix C for details.) However, it is possible to create efficient greedy approximate algorithms (Appendix D). We note that Chatterjee et al. [19] explored NP-hardness results and greedy heuristics for typo-tolerant password authentication, where the server returns true or false should the submitted password be a typo of the registered password. But, in our setting, MIGP oracle returns one of three possible answers. Therefore, their setting and results don't directly carry over to our setting.

In Appendix D, we present an efficient greedy algorithm for the $m = 0$ case, called GreedyMIGP. We now turn to measuring the efficacy of the greedy algorithm to understand the real-world threat of a malicious client attempting to extract data from the MIGP service. We assume the attacker has a guessing budget of $q \leq 10^3$. This setup assumes that the MIGP server will deploy some form of rate-limiting on queries from a client (as discussed in Section 7).

**Experiment setup.** For simulation, we assume the MIGP oracle is instantiated with $S_1$ data (see Section 4.2), and the attacker has access to only $S_2$. This simulates the situation where the attacker does not know the leaked data present in MIGP, and is trying to learn those breached passwords for a user. We sample 25,000 username, password pairs from $S_1$. For each pair, the attacker is given the username and required to find the target password. We compute the efficacy of an attack as the fraction of username, password pairs that the attacker can successfully guess. (As per our data division, none of the target usernames are present in $S_2$, and therefore the attacker cannot attempt a targeted credential tweaking-type attack.) We evaluate the security loss for 10 variants based on Das-R rules, and 100 variants based on wEdit rules. We first experiment with only server-side variant generation ($m = 0$); later in the section, we report the efficacy of breach extraction attacks when allowing client-side variant generation.

The $S_2$ dataset has 210 million passwords. If the attacker sets $W$ to all the passwords in $S_2$, it will make GreedyMIGP very slow to run (Fig. 15). We instead heuristically picked the top one million passwords as $W$ for the attack. These passwords are used by 24% of users in $S_2$.

For comparison, we also simulate a C3 service that does not provide checking for similar passwords, which we refer to as MIGP service with $n = 0$. For this case, the attack is simpler: query the MIGP service with the top $q$ passwords, and if any query returns match, output the queried password.

**Results.** The success probability of the attacker in guessing the target password using $q \in \{10, 100, 1000\}$ queries is shown in Fig. 8. We explain the $\beta$ values below; here we focus on the rows with $\beta = 0$. An attacker can learn 6.57% of passwords in less than a thousand guesses against an exact-checking C3 service ($n = 0, \beta = 0$). The attacker's success probability increases to 13.58% for $n = 10$ and 17.18% for $n = 100$. In the latter case, moving to a MIGP service may lead to a 2.8× increase in an attacker's ability to perform breach extraction attacks.

We observed a counter-intuitive pattern for $q = 10$ and 100: the attack success rate decreases with the increase of $n$ from 10 to 100. This is because large $n$ produces larger balls, making it easier to get in the ball, but harder to identify the correct password given a small query budget $q = 10$. Therefore, for small $q$, guessing the most weighted password ball may not be the optimal strategy. We plot the attack success for different values $q$ for $n = 10$ and $n = 100$ in Fig. 9. For query budget $q < 230$, increasing the number of password variants $n$ from 10 to 100 actually decreases this attack's success rate.

| β | n | q = 10 | q = 100 | q = 1000 |
|---|---|---|---|---|
| 0 | 0 | 1.64 (± 0.22) | 3.36 (± 0.35) | 6.57 (± 0.61) |
| | 10 | 2.14 (± 0.26) | 5.22 (± 0.56) | 13.58 (± 0.61) |
| | $10^2$ | 1.69 (± 0.17) | 3.54 (± 0.30) | 17.18 (± 0.73) |
| 10 | 0 | 0.03 (± 0.00) | 0.36 (± 0.08) | 2.80 (± 0.27) |
| | 10 | **1.19** (± 0.15) | 3.90 (± 0.40) | 12.12 (± 0.45) |
| | $10^2$ | **0.93** (± 0.11) | 2.43 (± 0.25) | 15.67 (± 0.61) |
| $10^2$ | 0 | 0.03 (± 0.03) | 0.37 (± 0.08) | 2.50 (± 0.30) |
| | 10 | 0.93 (± 0.13) | 2.71 (± 0.27) | 9.91 (± 0.42) |
| | $10^2$ | 0.79 (± 0.13) | 1.52 (± 0.26) | 10.91 (± 0.41) |
| $10^3$ | 0 | < 0.01 (± 0.00) | 0.18 (± 0.06) | 1.46 (± 0.14) |
| | 10 | 0.76 (± 0.11) | 1.42 (± 0.10) | 5.94 (± 0.16) |
| | $10^2$ | **0.72** (± 0.09) | 0.97 (± 0.11) | 9.21 (± 0.24) |
| $10^4$ | 0 | < 0.01 (± 0.02) | 0.03 (± 0.02) | 0.27 (± 0.03) |
| | 10 | 0.71 (± 0.10) | 1.02 (± 0.07) | 3.34 (± 0.23) |
| | $10^2$ | 0.70 (± 0.10) | 0.92 (± 0.11) | **4.87** (± 0.12) |

Figure 8: Attack success rate given different query budgets ($q$) for different attack scenarios. Here $n = 0$ (first row in each block) emulates existing exact checking C3 services. MIGP oracle uses Das-R and wEdit similarity rules for $n = 10$ and $n = 100$, respectively. The service blocks most frequent β passwords. All success rates are in percent (%) of 25,000 target users sampled from $S_1$. Standard deviations (shown in parenthesis) are measured across the 5 random folds of these pairs. Lower values imply better security.

**Blocklisting.** The abuse prevention mechanisms (e.g., slow hashing and API rate limits) used in current C3 systems can only slow down breach extraction attacks, but do not prevent them. We, therefore, propose a simple yet effective mechanism to reduce the attack success: blocklist the top β passwords so that an attacker learns nothing from the MIGP service should a user's password be one of them. The MIGP service can do so by removing all the blocklisted passwords and their variants from its breach database. (This will also reduce storage and bandwidth overhead as we show in Section 6.) These popular passwords are anyway unsafe to be used by any user irrespective of whether they are leaked or not. Therefore, a client application can warn the user who is using a password equal to or similar to one of the blocklisted passwords.

For our simulations, we assume the MIGP service blocks the β most frequent passwords according to $S_1$ and their variants (according to the setup). If an attacker queries the MIGP service with any of the blocklisted passwords the service always responds as none. Of course, the attacker is aware of the set of blocklisted passwords and their variants.

We experiment with different values of β as shown in Fig. 8; β = 0 denotes no blocklisting. Blocklisting reduces the attacker's success across all values of $n$. For $q \leq β$, the success probability of an attacker for $n = 10$ and $n = 100$ remains below that of existing C3 services (with $n = 0$ and β = 0), except for β = $q = 10^3$ in $n = 100$. We believe this is due to the higher ball size in case of $n = 100$. In this case, we need to blocklist β = $10^4$ passwords to reduce the attack success rate below existing C3 services. We highlight those numbers in the figure. As the top β passwords are blocklisted, the attacker can't learn if the user has a breached password that is one of the top β passwords. The attacker's best bet is to guess passwords that are outside the top β passwords. The password distribution follows Zipf's law [50], therefore leading to a significant decrease in breach extraction accuracy.

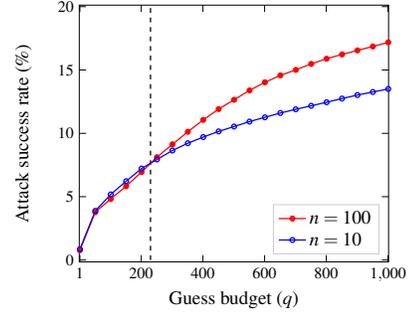

Figure 9: Comparison of attack success of breach extraction attack for Das-R ($n = 10$) and wEdit ($n = 100$). for different values of $q$. For query budgets $q \leq 230$ (black dashed line), success rates are slightly lower for the higher value of $n$.

We also compute the breach extraction success rate against users who do not use weak passwords. The results are shown in Fig. 16 (Appendix E). For these users, we found that the relative increase in attack success due to MIGP service is higher, but the absolute success rates are small. For example, for β = $10^4$ and $q = 10^3$, the attack success is 0.27% when $n = 0$, 2.98% when $n = 10$, and 2.56% when $n = 100$. Similar to Fig. 8, the attack efficacy for $n = 100$ is worse than that of $n = 10$ in most cases. We suspect that this is because our attack is not optimal, especially for a smaller number of guesses ($q \leq 10^3$).

**Security of client-side variant generation.** A client can generate $m$ variants of a password $w$ and check them all in parallel with the MIGP server. Due to the limitation of our MIGP protocol there is no way for the server to verify if the client has generated variants truthfully. Thus a malicious client can use this to expand its query budget by a factor of $m$: The client simply submits the next $m$ passwords computed using GreedyMIGP, to obtain in total $m \cdot q$ queries.

We show the success rate of this breach extraction attack for different $m$ and $n$ values with β = $10^4$ in Fig. 10. Allowing $m = 10^2$ variants on the client side increases the attack success by more than twice for any $q \leq 10^3$ compared to allowing only the server side $n = 10^2$ variants. Using the hybrid approach to variant generation with $m = 10$ and $n = 10$ reduces the attack success rate, but still remains significantly higher than $m = 0, n = 100$ setting. Therefore, we suggest that if the breach data is sensitive it is safer to disallow client-side variant generation and apply strict rate limiting.

| $n$ | $m$ | $q = 10$ | $q = 100$ | $q = 1000$ |
|---|---|---|---|---|
| $10^2$ | 0 | 0.70 ($\pm$ 0.10) | 1.10 ($\pm$ 0.05) | 4.87 ($\pm$ 0.12) |
| 10 | 10 | 1.02 ($\pm$ 0.07) | 3.34 ($\pm$ 0.23) | 8.49 ($\pm$ 0.09) |
| 0 | $10^2$ | 1.45 ($\pm$ 0.08) | 3.06 ($\pm$ 0.22) | 10.60 ($\pm$ 0.68) |

Figure 10: Comparing breach extraction attack success rate between generating $m$ variants on the client-side and $n$ variants on the server-side. Here we assume $\beta = 10^4$. The first row is the same as the last row in Fig. 8.

## 5.2 Security of the MIGP Protocol

Our MIGP protocol (given in Fig. 2) requires minimal changes to previously proposed [37] and currently deployed [47] protocols. This made deployment simpler, and also the cryptographic security is derived directly from the underlying protocol. Here we briefly summarize the security achieved by MIGP, considering in turn curious servers and malicious client threat models. MIGP communications must be protected with TLS, preventing any manipulations of the buckets or client queries by a network adversary.

As in the prior protocols, the MIGP server learns only the client's queried bucket ID. This reveals some bits of information about the username, but nothing about the queried password, assuming the password and username are independent. (Some users may choose passwords similar to their username, but it's unclear how a malicious server can usefully exploit this practice.) An actively malicious server can modify the result obtained by a client, e.g., by erroneously claiming passwords are not in the breach when, in fact, they are (or vice versa). This attack is possible also for deployed exact equality checking protocols [37, 47]. In theory, one could try to use techniques to prevent this, e.g., by having the server publish a commitment to the dataset and then performing zero-knowledge proofs of (non-)membership [39]. We do not believe this is necessary for breach alerting as such attacks would seem to have low value to attackers.

An encrypted bucket reveals to a client the number of entries in the bucket, and the updated entries and the time of updates to buckets will be revealed over time. This could conceivably have security implications in some contexts. As shown in Section 5.1, MIGP services are susceptible to breach extraction attacks, and therefore must employ different forms of rate limiting. Note that a malicious client can submit a query for a bucket for username $u$ but submit an OPRF request for username $u' \neq u$. This is true as well for existing C3 services. Thus rate-limiting should not be based (only) on bucket identifier, and instead on a client identifier (cookie or IP address) or a token mechanism such as PrivacyPass [23].

## 6 Performance Analysis

We implement a prototype of MIGP and conduct experiments to estimate its performance. We also compare it with existing C3 services, such as GPC [47] or IDB [37] (equivalent to MIGP with $n = 0$ and $m = 0$). We experiment with no blocklisting ($\beta = 0$) and blocklisting the $\beta = 10^4$ most frequent passwords (and their variants). We want to measure and compare: (1) storage overhead on the server side, (2) latency of running the protocol, and (3) total bandwidth usage. Here we use as breach dataset D the entire 1.14 billion username, password pairs from the dataset described in Section 4.2.

**Prototype implementation details.** We implement the MIGP client and server in Python 3.8, with petlib library for elliptic curve operations. We chose the elliptic curve group secp256k1 for $\mathbb{G}$ and set $\ell = 128$. For $H_1$, we use petlib's hash-to-point function to map the username, password pair to $\mathbb{G}$; internally it uses rejection sampling [30] with SHA256. We also use SHA256 for $H_2$. Should either of $H_1$ or $H_2$ be a slow hash, there will be additional overhead in precomputation. We select the most frequent $\beta$ passwords for blocklisting based on the dataset D. The server is built using the Flask [7] library and the client uses the requests [6] library to make queries. This prototype implementation is publicly available.[2]

For all the pre-processing of the data, we used a desktop with an Intel Core i9 processor and 128 GB RAM. We did not use any GPU to optimize hash computation in our experiments. For latency and bandwidth comparisons, we run the server (t2.medium) and client (t2.micro) on two different AWS EC2 instances running the Ubuntu 20.04 LTS image and located in two different regions — US-East and US-West.

**Precomputation overhead.** We precompute the buckets of PRF values for the entire breach dataset. MIGP, in comparison to exact-check C3 protocols, requires processing an additional $n$ variants for each leaked password and storing the resulting PRF values. The precomputation on the server involves computing $F_\kappa(x) = H_2(x, H_1(x)^\kappa)$, where $x = (u\|w)$ for the breached password and $F_\kappa(x) \oplus 1$, where $x = (u\|\tilde{w})$ for $\tilde{w} \in \tau_n(w)$. We use $H_2$ to reduce the representation size of the hash to 16 bytes, which saves disk space and bandwidth.

Generating $n = 10$ variants for a password using Das-R similarity rules takes less than 0.02 ms, whereas generating $n = 100$ variants using wEdit similarity rules takes 0.8 ms, on average. If we had used Argon2 as $H_2$, it would take 95 ms on average for computing the hash of one username, password pair, an estimated 361.5 CPU-years for pre-processing all username, password pairs and their variants on our reference implementation. Should breach data not be particularly sensitive, a deployment can skip slow hashing or use time-lock puzzles instead (see Appendix F).

The PRF values are then separated into buckets based on $H^{(l)}(u)$. Duplicate values could arise when pairs $(u, w_1)$ and $(u, w_2)$ satisfy the condition $\tau_n(w_1) \cap \tau_n(w_2) \neq \varnothing$, which can be common for users with credentials from multiple sites in the known breach. Duplicates should be either omitted (as we do in our prototype) or replaced with other variants. The former is better for performance, but note that the length of buckets now depends on relationships between different passwords — we are unsure whether this can be exploited by an attacker given the large number of users in each bucket.

---

[2] https://github.com/islamazhar/migp_python

| | w/o blocklisting | | w/ blocklisting | |
|---|---|---|---|---|
| $l$ | avg. | std. | avg. | std. |
| 16 | 1,751,666 | 36,832 | 1,431,876 | 30,107 |
| 20 | 109,479 | 9,192 | 89,492 | 7,513 |
| 24 | 6,842 | 2,297 | 5,592 | 1,877 |

Figure 11: Average bucket size of MIGP with $n = 100$ variants for each password on the leak dataset, which contains 1.14 billion unique username, password pairs.

The total storage cost for 1.14 billion unique username, password pairs and their variants would be 1.67 TB (considering every entry has $n = 100$ unique variants). Blocklisting reduces storage (and bandwidth) requirements, and we found that blocklisting the $10^4$ most popular passwords and their variants reduces the database size by 18% to 1.36 TB.

**Bucket size selection.** To make the private membership test protocols practical, C3 services use bucketing to partition the leaked dataset based on the prefix of the hash of the username. MIGP follows the same approach. However, as MIGP contains $n$ variants of each password, the buckets would be quite large for MIGP for the same number of buckets. Large bucket size will increase communication costs in terms of bandwidth and latency as the client has to download a larger amount of data. We can reduce the bucket size by increasing the number of buckets — by increasing the length of the hash prefix $l$. The average bucket sizes for different hash prefix lengths for MIGP are shown in Fig. 11. The bucket size, as expected, decreases exponentially with the prefix length ($l$). The average bucket size with blocking the most frequent $10^4$ passwords for $l = 16$ (which is used by GPC [47]) is 1.43 million, or 22.85 MB. Increasing the length of the bucket identifier $l$ to 20 reduces the bucket size to 1.37 MB.

**Latency & bandwidth comparison.** We measure and compare the latency and bandwidth requirements for running different compromised credential checking services: IDB-16 (also called GPC) [47], IDB-20 [37], WR19-Bloom [52], WR20-Cuckoo [53], and our protocols MIGP-Server, MIGP-Client, and MIGP-Hybrid. Although WR19-Bloom and WR20-Cuckoo were designed to check user's passwords in multiple web services, these protocols can be used for checking a user's leaked passwords. MIGP-Server, MIGP-Client, and MIGP-Hybrid are the different versions of our protocol with variants generated on the server side ($n = 10^2$), client side ($m = 10^2$), and both ($n = 10, m = 10$). IDB-16 and IDB-20 are implemented following the same construction as MIGP, but with different lengths of prefixes for bucketing and setting $m = n = 0$. For WR19-Bloom and WR20-Cuckoo, we use the corresponding authors' implementation[3] in Go but customize it for the client-server setting.

[3] https://github.com/k3coby/pmt-go

We pick 25 random passwords from the test data set T and run each C3 service protocol separately for different $n$, $m$ and prefix length $l$. We simulate the server data with dummy buckets containing $b$ entries, where value $b$ is randomly sampled from the normal distribution with the mean and standard deviation set to the values we computed in Fig. 11, with $\beta = 10^4$. The server and the client are executed in two different EC2 virtual machines located in two different availability zones in two coasts of the United States. They are connected via a 252 Mbits/sec network link.

We report the average latency with the breakdown for preparing the query, calling the API and waiting for the response, and finalizing the result for each protocol evaluation in Fig. 12. The overall time to prepare for a query takes less than 7 ms, for GPC, IDB, MIGP-Server and MIGP-Hybrid. The total computational cost for the server is very small compared to the client, however, the client spends time downloading the data from the server (leading to higher latency in MIGP compared to GPC and IDB). After the query, the client finalizes the result by computing $H_2$ of the username, password pair to compare with the bucket entries. Using slow hash function for rate limiting would add about 95 ms to the query preparation to all protocols. MIGP-Client takes 100x more time in query preparation due to generating the variants and checking them. MIGP-Client can be particularly expensive with rate limiting using slow hashes, such as Argon2. It can take more than 10 seconds for a complete run of the protocol run. MIGP-Hybrid strikes a balance between server storage and query preparation time. It reduces the storage cost and query time by a factor of 10 compared to MIGP-Server and MIGP-Client, respectively.

The slowest among all, is WR19-Bloom and WR20-Cuckoo protocols, taking more than 38 sec for one complete protocol execution. The primary contributors to the latency are: (a) before a query, the client has to encrypt each entry of the Bloom filter homomorphically (using Paillier encryption [40]), (b) the client has to send all the encrypted Bloom filter entries to the server, which is quite large (216.8 MB), and (c) the server has to compute large group multiplications over *all* entries in the Bloom filter. WR20-Cuckoo protocol uses Cuckoo hashing [27] which improves overall latency and bandwidth, however, still falls short of being practical due to high computational overhead on the server.

MIGP-Server (with $n = 10^2$ and $l = 20$), MIGP-Client (with $m = 10^2$ and $l = 16$) and MIGP-Hybrid (with $n = 10, m = 10$ and $l = 20$) takes less than 534 ms to compute a query if we don't use rate limiting, which is comparable to currently-in-use IDB-16 (with $l = 16$) protocol. The overhead for MIGP-Server stems from the high bandwidth usage due to large bucket sizes. The buckets can be cached on the client-side or served directly from CDNs (such as Cloudflare) as practiced by HIBP [14] to improve performance. Client-side caching of buckets saves fetching the same bucket again for checking different passwords for the same username. As most users have only a few email addresses, this can save significant network bandwidth and time over multiple queries.

|  |  |  |  | Client side latency (ms) | | | | |
| --- | --- | --- | --- | --- | --- | --- | --- | --- |
| C3 service | $l$ | Server storage | B/w (MB) | Query prep. w/o rate limit | Query prep. w/ rate limit | API call | Fina-lize | Total w/o rate limit | Total w/ rate limit |
| IDB-16 (GPC [47]) | 16 | 15 GB | 0.23 | < 1 | 96 | 321 | < 1 | 322 | 417 |
| IDB-20 [37] | 20 | 15 GB | 0.01 | < 1 | 96 | 125 | < 1 | 126 | 221 |
| WR19-Bloom [52] | 20 | 1.0 TB | 177.20 | 6,990 | 7,065 | 39,045 | < 1 | 46,035 | 46,110 |
| WR20-Cuckoo [53] | 20 | 0.8 TB | 0.81 | 59 | 155 | 38,560 | < 1 | 38,619 | 38,715 |
| MIGP-Server | 20 | 1.5 TB | 1.43 | < 1 | 95 | 498 | 3 | 501 | 596 |
| MIGP-Client | 16 | 15 GB | 0.23 | 46 | 10,713 | 450 | 38 | 534 | 11,202 |
| MIGP-Client | 20 | 15 GB | 0.02 | 48 | 10,007 | 390 | 38 | 476 | 10,435 |
| MIGP-Hybrid | 20 | 0.2 TB | 1.43 | 7 | 953 | 421 | 12 | 440 | 1,386 |

Figure 12: Average latency (in milliseconds) for checking one password via different private membership or similarity test protocols used in different C3 services. with different parameters. IDB-16 and IDB-20 do not use any variants. MIGP-Server and MIGP-Client generate $10^2$ variants on the server and the client side, respectively. WR19-Bloom uses Bloom filter to reduce the b/w requirement. For rate limiting we use Argon2 as $H_2$, which takes around 95ms to compute. All latency measurements are averaged over 25 complete API calls with standard deviations < 10%.

## 7 Deployment Discussion

We worked with Cloudflare, a major CDN and security company [12], to deploy the MIGP protocol (1) as a public-facing API similar to HIBP, and (2) as a new breach alerting feature within Cloudflare's web application firewall (WAF) product [9]. MIGP is deployed as an opt-in feature in WAF, which detects login requests to Cloudflare customer websites, extracts username and password fields, and queries a MIGP service deployed on Cloudflare Workers [10]. The result of the MIGP query is added to an HTTP header that is forwarded to the customer login service, informing them should the login request be utilizing a breached credential or ones similar to them. The libraries underlying the MIGP implementation have been open-sourced and are publicly available [8]. In this section, we present some deployment considerations and the lessons we learned.

**Deployment details.** During pre-processing, the breach database is transformed into MIGP buckets. We post-process the OPRF outputs using HKDF [35] to generate a 21-byte hash value; the last byte is XORed with a one-byte flag denoting whether a bucket entry is an exact match or a variant. Slow hashing is supported by the implementation, but applying slow hashing at the scale is expensive and our initial deployment omits it. We discuss this more below.

The buckets of credentials produced in the pre-processing step are stored in Workers KV [11], a high-performance, distributed key-value store. Our deployment uses 20-bit bucket prefixes with $n = 8$ variants per entry generated using the Das-R rules and without blocklisting popular passwords ($\beta = 0$). The deployment caps each individual bucket size to 25 MB, which under this configuration should support breach data up to 64 billion entries. The MIGP service is able to serve over 50% of client requests in under 135ms, and 95% of requests in under 573ms. Most performance overhead is due to the cost of fetching buckets form Workers KV store, as only frequently accessed buckets are cached at 250 datacenters that are running Workers. Other buckets must be fetched from a centralized data store which adds latency.

Our current implementation does not support client-side variants. As shown earlier (Fig. 12), enabling client-side variant generation in the future may provide attractive performance benefits. This must be balanced against the risk of breach extraction attacks (see Fig. 10).

**Breach extraction attacks.** A key concern as we designed and discussed MIGP deployment at scale was gauging the risk of breach extraction attacks. Any client can attempt to mount such an attack against the public API. The WAF deployment does not necessarily provide malicious web clients with a MIGP oracle: the result of MIGP queries are only shared with the login service and not the client. Login services should not reveal MIGP outputs to unauthenticated clients.

For both deployments we have thus far only utilized datasets that are widely available on underground forums, obviating the concern about breach extraction attacks in the short term. To use more sensitive breaches in the future, further mitigations will need to be enabled, including popular password blocklisting and rate limiting. Our deployment already benefits from rate-limiting of individual IP addresses and other anti-automation techniques [15]. We note that a common rate-limiting approach is to require clients to obtain an API key through some slow (or paid) registration process, but this approach won't work for WAF deployment scenario.

Another rate-limiting approach would be to use slow hashing. Recall that the MIGP protocol uses two hash functions within the OPRF, computing outputs as $H_2(u\|w, H_1(u\|w)^\kappa)$. Either of the hashes can be made computationally expensive to both slow down online breach extraction attacks and to make offline hash cracking attacks harder should the

MIGP server be compromised. There are nuanced security-computation trade-offs between the choice of which to make slow. If $H_2$ is expensive the client cannot do offline processing of the slow hash without communicating with the server, which is not true if only $H_1$ is slow. However, one benefit of having just $H_1$ slow is that the server can store the intermediate $H_1(u\|w)^\kappa$ values for faster key updates (see below). Google Password Checkup (GPC) [47] uses a slow hash for $H_1$ and a fast hash for $H_2$.

An alternative approach to slow hashing is to use asymmetric hashing, also called proof-of-work [24, 31] or time-locked puzzles [45]. We discuss these approaches in more detail in Appendix F.

**Bucket updates.** MIGP services (like other C3 services) may periodically update their leaked data, such as when new breaches are exposed online. This will require adding new credentials to the buckets. Updating the buckets with OPRF outputs under the same key could, in principle, allow an attacker to identify the newly added username, password pairs. Although it is unclear how this leakage can be exploited, it would be better to avoid the leakage entirely. One way is to rotate the OPRF key $\kappa$ every time there is a new leak. However, recomputing the OPRF output from the stored breach data will be computationally very expensive given the slow hash function. Assuming $H_1$ is slow, an optimization would be to have the server record the output of the group multiplication $H_1(u\|w)^\kappa$ in some offline, safe storage. Then the new OPRF outputs can be computed for the new key $\kappa'$ by raising $H_1(u\|w)^\kappa$ values to $\kappa'/\kappa$, and applying the fast hash $H_2(\cdot)$ to them. This approach is similar to the key rotation mechanism used by Pythia [26].

**MIGP warnings: effectiveness and usability.** To estimate the effectiveness of the MIGP service, we instrumented the WAF deployment to measure the ratio of the number of login attempts that MIGP flagged as similar to the number that MIGP flagged as an exact match. The average ratio over the period of a week is 0.2 (with 0.01 standard error of the mean), implying that MIGP flags 20% more login attempts compared to an exact-checking C3 system. This represents a significant improvement by MIGP over exact-checking in terms of alerting on credentials that are vulnerable to attacks such as those based on pass2path [41]. Our instrumentation does not record how many WAF-monitored attempts correspond to vulnerable accounts (e.g., attempts will include some number of incorrect submissions and attacks), but customer services can distinguish between these cases and act appropriately.

Prior work has shown that users may not be responsive to breach alerts [44]. We expect that MIGP deployments will face a similar challenge. Server-side breach alerting, like our WAF deployment, allow high-security services to force users to change MIGP-flagged passwords. One open question prompted by our work is how best to communicate to users that their password is similar to a breached password and how to guide them towards safer choices.

# 8 Conclusion

In this work, we tackled the problem of building MIGP, an updated version of C3 systems that can securely warn users from selecting passwords similar to (and same as) a breached password which can be vulnerable to credential tweaking attacks. Via comparing different similarity metrics we show that computing variants of the password using weighted edit distance rules provide the best combination of performance and efficacy. Underlying MIGP is a secure private similarity test (PST) protocol. Despite secure PST, MIGP protocols can still be vulnerable to breach extraction attacks, where an attacker can extract leaked (but not yet public) credentials from a MIGP service. We show that the attacker's success probability can be reduced significantly using blocklisting popular passwords. We implement and show that MIGP achieves computational overhead comparable to C3 services. Finally, we deploy MIGP with Cloudflare and provide nuanced discussions about deploying MIGP in practice.

# Acknowledgements

We thank the anonymous reviewers for their insightful comments and suggestions, as well as Adam Oest for his work as shepherd for the paper. This work was supported in part by NSF grant CNS-2055169 and the Wisconsin Alumni Research Foundation.

## A  Assessing performance feasibility

We compare the performance of the top four similarity measures — P2P, Das-R, wEdit and PPSM — to understand the feasibility of their deployment as a C3 service. The three generative algorithms perform a PMT with the breached passwords and their variants generated on the server. PPSM computes similarity by mapping the passwords to 100-dimensional vectors and comparing their dot product to a threshold. The resultant list of boolean is summed and sent to the client. Therefore, the implementation of PPSM-based MIGP doesn't require the generation and storage of similar passwords, but involves computing private dot products, comparisons, and summation.

To estimate the cost of performing similarity matching, we use a bucket of 10,000 username, password pairs (without any variants). We use $n = 100$ for P2P, Das-R, and wEdit, and $\theta = 0.83$ for PPSM. The OPRF based PMT protocol is implemented in Python and uses secp256k1 elliptic curve implemented in petlib [5]. We implement the PPSM based protocol using Crypten [34]. Timing experiments were performed on a machine with an Intel Core i9 processor and 128 GB RAM, and here we run the entire protocol within the same machine (without network overhead). P2P uses an Nvidia GTX 1080 GPU along with the processor to run the pass2path neural network for pre-processing.

We summarize the results in Fig. 13. PPSM-based approach to MIGP takes 16 seconds to complete a query, while all other approaches take < 1 second. Note that these measurements, which do not include network latency, should be considered lower bounds on performance. Crypten uses secret-sharing to execute the MPC protocols, therefore requires more than one round trip. The columns of Fig. 13 are ordered from left to right in decreasing order of our perception of how critical this aspect of the protocols is to deployment. As PPSM is slower than other approaches for executing a query, we focus on the generative methods. We leave as an open question whether one can make another 2PC-based protocol fast enough for reasonably sized buckets.

| Similarity measure | Latency | B/w | Compat. | Storage per bucket | Precomp. |
|---|---|---|---|---|---|
| wEdit ($n=100$) | < 1 sec | 14 MB | Yes | 14 MB | 41 sec |
| P2P ($n=100$) | < 1 sec | 14 MB | Yes | 14 MB | 180 sec |
| Das-R ($n=100$) | < 1 sec | 14 MB | Yes | 14 MB | 0.5 sec |
| PPSM ($\theta=0.83$) | 16 sec | 1.6 KB | No | 8 MB | 1 sec |

Figure 13: Performance (latency, bandwidth, storage, etc.) summary of different similarity measures. All the numbers are based on a bucket of size $10^4$. The trade-offs are also ranked left to right based on the importance to deployment. Here latency does not include n/w or i/o cost.

| Das-R | wEdit | Rule | (%) of matches |
|---|---|---|---|
| 1 | 1 | Del last char | 27.7 |
| 2 | 3 | Switch $1^{st}$ char case | 20.6 |
| 3 | 2 | Del last 2 char | 15.2 |
| 6 | 4 | Ins '1' at end | 13.4 |
| - | 6 | Ins 'Caps' at beg | 7.6 |
| 4 | 5 | Del last 3 char | 6.6 |
| 9 | 7 | Del $1^{st}$ char | 4.7 |
| 5 | - | Ins '0' at beg | 1.5 |
| - | 8 | Subs '1' at end | 1.0 |
| 10 | - | Ins '0' at end | 0.7 |
| - | 10 | Ins '123' at end | 0.5 |
| 7 | 9 | Ins 'a' at beg | 0.4 |
| 8 | - | Ins 'q' at beg | 0.1 |

Figure 14: Rules for generating password variants and the % of password pairs matched by the rule among 9,141 vulnerable pairs found in a randomly sampled $10^5$ password pairs. We also show their ranks according to Das-R and wEdit.

## B  Rules-based similar passwords generation

We used three rule-based approaches for generating similar passwords: Das [22], a reordered variant of Das which we call Das-R, and wEdit. The top-performing edit rules based on our dataset $S_1$ are shown in Fig. 14. We also report the percentage of vulnerable password pairs in T explained by each rule. Deleting characters towards the end, and adding SHIFT or CAPS LOCK at the beginning are the most common rules that users use to modify their passwords. While the top rules capture the common transformations, it fails for subtle edits that are otherwise guessed by pass2path. Some example of such pairs are: ('20041981', '200481'), ('thingsome', 'thing.some'), ('nikaprudova', 'nika_prudova'), ('MADRE000', 'padre000'), ('jessiemax1', 'jessie1').

## C  Optimal breach extraction attack is hard

Let $\mathcal{W}$ be a set of all possible strings up to some length (say, 30), and $W \subseteq \mathcal{W}$ be the set containing all possible password strings for users $U$, with an associated probability distribution $p$ of being chosen by a user. Recall we defined $\tau : W \to 2^{\mathcal{W}}$ such that $\tau(w)$ is the set of passwords similar to $w$. (We show how to instantiate $\tau$ in Section 4.) We assume $w \notin \tau(w)$, and $n = \max_w |\tau(w)|$.

The MIGPGuess adversary (in Fig. 7) tries to guess the exact target password $w$ given access to the MIGP$(\cdot)$ oracle. We modify the game to measure the advantage of an attacker as the expected number of queries to the respective oracles to guess a password, without any limit on the query budget. The oracles keep track of the total number of queries.

The advantage of the MIGPGuess adversary $\mathcal{A}$ is defined as the expected guess rank: $G^{\mathsf{MIGP}}(\mathcal{A}) = \mathbb{E}[\mathsf{MIGPGuess}(\mathcal{A})]$.

An MIGPGuess-adversary $\mathcal{A}^*$ is optimal if for all $\mathcal{A}$ it holds that $G^{\mathsf{MIGP}}(\mathcal{A}^*) \leq G^{\mathsf{MIGP}}(\mathcal{A})$. The optimal adversary $\mathcal{A}^*$ builds a ternary decision tree to query MIGP such that the expected guess rank is minimized. We show that building such a decision tree that minimizes the guesswork is NP-hard, and so does the optimal attack against MIGP.

**Definition 1** (OMIGPGuess). *Given $(W, p, \tau)$, we define optimal MIGP guess (OMIGP) problem as building the query tree for $\mathcal{A}^*$ that minimizes the expected guess rank for distribution $p$ over $W$ with MIGP similarity measure being $\tau$.*

**Theorem C.1.** OMIGPGuess *problem is NP-hard.*

**Proof:** To prove this theorem, one might be tempted to reuse the result from Chatterjee et al. [19, 21], who investigated guessing attacks against a server that allows the user to login with a small set of typos in the context of typo-tolerant password checking. Although the setting is similar, there is one crucial difference: MIGP reveals whether the query is a match or is similar to a password, but the password typo correction oracle does not reveal whether the password was an exact or near match. This seemingly minor distinction implies we can't use their technique.

We, therefore, show that OMIGPGuess problem is NP-hard by reducing the optimal binary decision tree (OBDT) problem to OMIGPGuess in polynomial-time. Because OBDT does not have a polynomial-time solution [36], OMIGPGuess also cannot have a polynomial-time solution.

**Binary decision tree (BDT).** Given a set of $n$ items $X = \{x_1, x_2, \ldots, x_n\}$ with associated probability distribution $p_X$ and a set of $m$ questions (functions) $Q$ such that $Q_i : X \mapsto \{0, 1\}$, the goal is to find a decision tree where the questions $q \in Q$ are specified in all the internal nodes while the items $x_i$ fill the root nodes of the tree. The expected depth of the tree is defined as $\sum_{x \in X} p_X(x) \cdot d(x)$, where $d(x)$ is the depth — distance from the root — of the element $x$ in the binary tree.

**Definition 2** (Optimal BDT problem (OBDT)). *Given $(X, p_X, Q)$, the problem is to build a binary decision tree that has the least expected depth, $\sum_{x \in X} p_X(x) \cdot d(x)$.*

Laurent and Rivest have shown OBDT problem is NP-hard [36]. We show that OMIGPGuess is NP-hard by giving a polynomial-time reduction of an arbitrary instance of OBDT problem to OMIGPGuess. Thus if there exists a polynomial-time solution to OMIGPGuess, then we can solve OBDT in polynomial-time as well, which is a contradiction.

An instance of OMIGPGuess problem is defined as $(W, p, \tau)$, where $W$ are the set of strings, $p$ is a probability distribution over $W$, and $\tau$ is a similarity measure. Given an instance of BDT problem $(X, p_X, Q)$, we can construct an instance of OMIGPGuess problem as follows. For this we set $W = X \cup Q$ (assuming elements in $Q$ are distinct from $X$); $p(w) = p_X(w)$ if $w \in X$, and 0 otherwise; and $\tau(w) = \{y \mid Q_w(y) = 1\}$ if $w \in Q$, and $\varnothing$ otherwise.

```
GreedyMIGP(W, p, B, q):
W' ← ⋃_{w∈W} τ(w) ∪ {w}
for i ← 1 to q do
    w̃_i ← argmax_{w̃∈W'} p(B(w̃)) ;  r ← MIGP(w̃_i)
    if r = similar then
        for w ∈ W\B(w̃_i) do  p(w) ← 0
        W' ← ⋃_{w∈B(w̃_i)} τ(w) ∪ {w}
    else if r = none then
        W ← W\B(w̃_i) ;  for w ∈ B(w̃_i) do  p(w) ← 0
    else if r = match then  return w̃_i
    W' ← W' \ {w̃_i}
return argmax_{w∈W} p(w)
```

Figure 15: Greedy algorithm for finding $q$ guesses to MIGP oracle (Fig. 7). Here $B : \mathcal{W} \mapsto 2^W$ is a function such that $B(\tilde{w}) = \{w \in W \mid \tilde{w} \in \tau(w)\}$.

Let $T$ be the ternary decision tree for OMIGPGuess problem, where each node has three children for each type of MIGP output. The leaf nodes of the tree are the passwords in $W$. The distance of a password $w$ from the root is the number of queries it take to guess $w$, which we denote as $d(w)$ here. As $T$ is optimal, the guesswork $\sum_{w \in W} p(w) \cdot d(w)$ is minimum. Also note that, because $p(w) = 0$ if $w \in Q$ (as per the reduction above), $\sum_{w \in X} p(w) \cdot d(w)$ is minimum. This is the same as the property of OBDT. Therefore, we can build the required binary decision tree $T'$ by removing the edges for the exact match of the questions (where $w \in Q$).

Thus, we show that one can reduce an instance of OBDT problem into an instance of OMIGPGuess problem, and the solution of OMIGPGuess will provide a solution to OBDT. This contradicts the fact that OBDT is NP-hard, therefore, OMIGPGuess cannot have a polynomial-time solution. This concludes the proof. ∎

The BDT problem will have a unique solution only if $m \geq \log_2 n$ and no two objects have the same output for all the questions. Since BDT reduces to OMIGPGuess, the same conditions apply for OMIGPGuess as well.

## D  Greedy approximation of OMIGPGuess

Finding an optimal guessing strategy that minimizes the expected guess rank is NP-hard. However, attackers could still find approximate solutions that minimize the expected guess rank. We present a greedy algorithm GreedyMIGP Fig. 15. We define the *ball* $B(\cdot)$ of a variant $\tilde{w} \in \mathcal{W}$ as the set of passwords that share a common variant. That is, $B(\tilde{w}) = \{w \in W \mid \tilde{w} \in \tau(w)\}$. The probability of a ball $p(B(\tilde{w}))$, also called the weight of a ball, is the sum of the probabilities of the passwords in the ball.

The attacker begins with a set of potential passwords $W$ of the target user. In iteration $i$, the attacker picks the guess $\tilde{w}_i$ that has the highest ball weight, and based on the response from the MIGP oracle, it updates the set of potential passwords. In particular, if the response is none, then it removes all the passwords in $B(\tilde{w}_i)$ from $W$. If the response is similar, then it knows that the target password is one of the passwords in $B(\tilde{w}_i)$, and so it sets the probability of all other passwords to zero and limits the search to the passwords in $B(\tilde{w}_i)$ and their variants. It is important to leave the variants in because they may make the ball heavier than when the ball was centered on passwords from only $B(\tilde{w}_i)$. If the response is match, then it stops and outputs the guess $\tilde{w}_i$ (and wins the game).

| β | n | q = 10 | q = 100 | q = 1000 |
|---|---|---|---|---|
| 10 | 0 | 1.21 (± 0.17) | 2.48 (± 0.25) | 5.64 (± 0.51) |
|  | 10 | **0.14** (± 0.06) | **1.85** (± 0.15) | 9.50 (± 0.31) |
|  | $10^2$ | **0.03** (± 0.02) | **0.44** (± 0.08) | 10.65 (± 0.16) |
| $10^2$ | 0 | 0.78 (± 0.10) | 1.40 (± 0.17) | 2.49 (± 0.30) |
|  | 10 | **0.12** (± 0.02) | 1.78 (± 0.28) | 8.57 (± 0.47) |
|  | $10^2$ | **0.03** (± 0.03) | **0.67** (± 0.13) | 6.46 (± 0.27) |
| $10^3$ | 0 | 0.72 (± 0.07) | 0.89 (± 0.08) | 1.46 (± 0.14) |
|  | 10 | **0.23** (± 0.02) | **0.78** (± 0.12) | 5.61 (± 0.54) |
|  | $10^2$ | **0.04** (± 0.01) | **0.09** (± 0.03) | 2.07 (± 0.27) |
| $10^4$ | 0 | < 0.01 (± < 0.01) | 0.03 (± 0.03) | 0.27 (± 0.03) |
|  | 10 | **<0.01** (± < 0.01) | 0.41 (± 0.07) | 2.98 (± 0.25) |
|  | $10^2$ | 0.00 (± 0.00) | 0.26 (± 0.21) | 2.56 (± 0.20) |

Figure 16: Breach extraction attack success when the target password is not one of the blocked passwords or their variants.

The greedy algorithm is not optimal but provides a good approximation of the success of the optimal attacker. Whether an efficient algorithm with tighter approximation bounds exists remains an open question. Following the seminal work of Chakaravarthy et al. [18]), we find the expected guesswork due to the greedy algorithm GreedyMIGP can be as high as $O(\log |W|)$ factor of the minimum expected guesswork $G^{\text{MIGP}}(\mathcal{A}^*)$. This approximation factor is quite large, especially when $|W|$ is very large. Nevertheless, this shows that it is possible to compute approximate solutions that might help an attacker guess a user's leaked password stored in MIGP server more effectively (than existing C3 services).

The complexity of the algorithm is $O(qn^2|W|)$. The attacker can decide on $W$ that they believe will likely contain the target password, e.g., popular passwords from prior public password breaches.

## E  Breach extraction attack (contd.)

To understand the security impact on users who uses strong password — passwords that are not blocked by MIGP, we sample of 25,000 username, password pairs randomly from T such that the target password is not in the blocklisted set, for $\beta > 0$. We show the results of greedy breach extraction attack GreedyMIGP in Fig. 16. Notably, for $q \leq 100$, our greedy attack performs worse than that against exact checking C3 service (shown in bold), and the attack success rate is worse for $n = 100$ compared to $n = 10$. We saw the similar trend in Fig. 8 as well. We believe this is because our greedy algorithm is sub-optimal, especially for smaller number of guesses and future work should explore other heuristic attacks for a smaller number of guesses.

## F  Rate-Limiting Client Queries

As shown in Section 5.1, to reduce the effect of the breach extraction attacks, MIGP must limit access to the service. Cryptographic rate-limiting ensures the client performs significantly more work than the server to make a query.

MIGP can use a slow, computationally expensive hash function such as Argon2 [1] or Scrypt [43] for $H_2$ (or $H_1$, see the trade offs in Section 7) in the PRF $F_\kappa$. For example, computing a slow Argon2 hash with default parameters [3] on a desktop with Intel Core i9 processor and 128 GB RAM takes about 97 ms. However, this also requires the server to compute the slow hash during pre-processing. We estimate that computing $F_\kappa$ for 1.14 billion unique username, password pairs and their $n = 100$ variants will require approximately 361.5 CPU-years of computational power.

An alternative approach would be to use a time-lock puzzle [38, 45] to slow down client queries to MIGP. Time-lock puzzles, first introduced by Rivest et al. [45], are a type of verifiable delay function (VDF) [16], where knowledge of trapdoor information makes computing a hash function significantly faster. Following the construction in [45], we can set $H_2$ to be computed as follows. The MIGP server computes a large RSA modulus $N = pq$, where $p$ and $q$ are two large randomly chosen secret primes. Let $\nu$ be the cost factor and $H_2(x) = \text{SHA256}(x)^{2^\nu} \mod N$, for any binary string $x \in \{0,1\}^*$. The server can compute $H_2$ efficiently as $H_2(x) = \text{SHA256}(x)^{2^\nu \mod \phi(N)} \mod N$, where $\phi(N) = (p-1) \cdot (q-1)$. The time complexity of such an operation would be bounded by the size of $N$ in bits. While for the client, which will not know the factors of $N$, computation of $H_2$ will need to perform $\nu$ squaring modulo $N$ sequentially (each time squaring the prior result). By setting the value of $\nu$ accordingly the server can increase the computational cost. The advantage of using time lock puzzles is that repeated squaring is an intrinsically sequential process and can't be parallelized. We estimate that the server would need 0.53 ms to set up a time-lock puzzle that would take 100 ms to solve for the client. This corresponds to approximately 1.8 CPU-years of computational power to finish computing $H_2$ of 1.14 billion unique username-password pairs and their $n = 100$ variations.

A third alternative approach to throttle client queries is to add a small secret value to the hash function $H_2$, $H_2(x) = \text{SHA256}(x\|r)$, where $r$ is randomly chosen from $\{0,1\}^\nu$ for each username, password pair. The server does not store $r$ (or share it with the client). Therefore, the client has to brute-force the value of $r$. For example, assuming a (malicious) client can do 10 million SHA256 hashes per second, the server can set the value of $\nu$ to be 21 bits, which will in expectation ensure 100 ms client-side computing cost. The server will require approximately 3.1 CPU-hour for precomputation. One drawback of this approach is that the client can parallelize the computation of hashes, and it does not guarantee $2^\nu$ sequential operations, unlike time-lock puzzles.